\documentclass[fleqn,twoside]{article}
\usepackage{espcrc2}

\usepackage{graphicx}
\usepackage[figuresright]{rotating}

\newcommand{\AmS}{{\protect\the\textfont2
  A\kern-.1667em\lower.5ex\hbox{M}\kern-.125emS}}

\title{Scalar field dark energy and Cosmic Microwave Background}

\author{C. Baccigalupi
\address[SISSA]{SISSA/ISAS, Via Beirut 4, 34014, Trieste, Italy and 
Lawrence Berkeley National Laboratory,
1 Cyclotron Road Mailstop 50-205, Berkeley, CA 94720}
\thanks{bacci@sissa.it},
A. Balbi, 
\address[ROMEII]{Dipartimento di Fisica, Universit\'a di Roma ``Tor 
Vergata" and INFN, Sezione di Roma II, Via della Ricerca Scientifica 
1, 00133 Roma, Italy}
\thanks{balbi@roma2.infn.it}
S. Matarrese 
\address[PADUAUNI]{Dipartimento di Fisica ``Galileo Galilei", 
Universit\'a di Padova and INFN, Sezione di Padova, via Marzolo 8, 
35131 Padova, Italy}
\thanks{matarrese@pd.infn.it}
F. Perrotta
\address[PADUAOBS]{Lawrence Berkeley National Laboratory,
1 Cyclotron Road Mailstop 50-205, Berkeley, CA 94720 and 
Osservatorio Astronomico di Padova, Vicolo dell'Osservatorio 
5, 35122 Padova, Italy}
\thanks{perrotta@materia.lbl.gov}
N. Vittorio \addressmark[ROMEII]
\thanks{vittorio@roma2.infn.it}
}

\begin{document}

\begin{abstract}
A dynamical scalar field represents the simplest generalization 
of a pure Cosmological Constant as a candidate to explain the 
recent evidence in favour of the accelerated cosmic expansion. 
We review the dynamical properties of such a component, and argue 
that, even if the background expectation value of this field is 
fixed and the equation of state is the same as a Cosmological 
Constant, scalar field fluctuations can still be used to distinguish 
the two components. We compare predicted spectra of Cosmic Microvave 
Background (CMB) anisotropies in tracking scalar field cosmologies 
with the present CMB data, in order to get constraints on the 
amount and equation of state of dark energy. 
High precision experiments like SNAP, 
{\sc Planck} and {\sc SNfactory}, together with the data on 
Large Scale Structure, are needed to probe this issue with the necessary 
accuracy. Here we show the intriguing result that, with a strong prior 
on the value of the Hubble constant today, the assumption of a flat 
universe, and consistency relations between amplitude and spectral 
index of primordial gravitational waves, the present CMB data at 
$1\sigma$ give indication of a dark energy equation of state larger 
than -1, while the ordinary Cosmological Constant is recovered at 
$2\sigma$. 
\vspace{1pc}
\end{abstract}

\maketitle

\section{Introduction}
\label{introduction}

One of the most exciting surprises in modern cosmology is the 
evidence of an acceleration in the cosmic expansion 
from type Ia supernovae \cite{PERL,RIESS}. To explain this 
result, about $70\%$ of the present cosmological energy density 
should be some sort of vacuum energy having a negative equation 
of state, like a Cosmological Constant, already considered by A. Einstein. 
However, the theoretical difficulties due to the too low 
value of the vacuum energy today with respect to the early universe 
allow, if not require, the introduction 
of a more general concept which is now widely known as ``dark energy". 
The first proposal in this sense was made well before the 
evidence of cosmic acceleration, already in 1988, by 
C. Wetterich, P. Ratra and P.J.E. Peebles \cite{FIRSTDE}, 
replacing a Cosmological Constant into the Einstein equations 
with a dynamical scalar field $\phi$, also known as ``Quintessence". 
Even if the system gets obviously more complicated, for 
general forms of the potential $V(\phi )$ there exist attractor 
trajectories for the evolution of the background expectation 
value of $\phi$. These trajectories have been proposed to alleviate, 
at least classically, the fine-tuning required in the early universe, 
when the typical energy scales were presumably Planckian, 
$10^{76}$ GeV$^{4}$, to generate a vanishing relic vacuum energy as 
it is observed today, at the level of $10^{-49}$ GeV$^{4}$ 
\cite{TRACKSCALE}. However, these scenarios do not solve 
the coincidence problem, merely why we are living in the epoch 
in which dark energy and matter have roughly the same energy density. 
Even if some attempts to address this issue exist in scalar field 
dark energy scenarios \cite{COINCIDENCE}, it remains largely 
a mystery. \\
Quintessence models in scalar tensor cosmologies are widely known 
as ``Extended Quintessence" scenarios \cite{EQ,TEQ}, where the idea 
is to connect dark energy to Gravity through a non-minimal coupling 
with the Ricci scalar in the fundamental Lagrangian. 
\begin{figure}
\includegraphics[height=3.in,width=3.in]{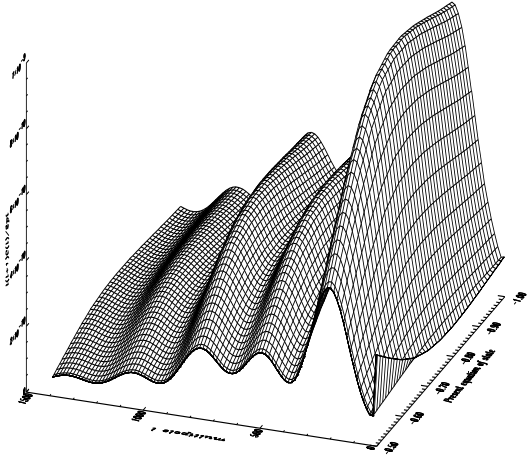}
\caption{Effects of scalar field dark energy 
models on CMB anisotropy: $w_{Q}=-1$ corresponds 
to a cosmological constant; projection and ISW effects 
are evident for $w_{Q}>-1$.}
\label{f1}
\end{figure}
These models have been explored under 
different perspectives in literature \cite{EQWORKS}, including 
recent important consequences on the structure formation process 
\cite{EQC}. Possible couplings with the matter sector of the Lagrangian 
have been also studied \cite{CQ}. In addition, intriguing models 
of dark energy, not based on scalar fields, have been proposed 
\cite{DEPROPOSALS}.\\ 
The evidence in favour of dark energy comes, in addition to 
type Ia Supernovae, from Cosmic Microwave Background (CMB) and 
Large Scale Structure (LSS) \cite{LAMBDAEVIDENCES}. 
Recently several authors tried to constrain the dark energy equation 
of state $w_{Q}$ \cite{BRAX,AME,OUR1,OUR2,MELCH,SVED}. Even if 
interesting results can be obtained assuming reasonable values 
of some parameters \cite{OUR2}, a true measure of the dark energy 
equation of state requires the precision of planned experiments 
like SNAP \cite{SNAP} and {\sc SNfactory} \cite{SNfactory} 
together with observations of LSS \cite{LSSSURVEYS} 
and of the CMB with MAP \cite{MAP} and {\sc Planck} \cite{PLANCK}.\\ 
In Section 2 we focus on the role of fluctuations in scalar 
field dark energy, pointing out that even in the case of a 
constant background expectation value for the scalar field 
$\phi$, fluctuations can still be present and are ultimately the 
signature of a dynamical component instead of a pure 
Cosmological Constant. 
In Section 3 we show interesting results obtained by 
comparing CMB spectra predicted in tracking minimally coupled 
Quintessence scenarios with available data \cite{OUR2}. 

\section{Scalar field dark energy fluctuations}
\label{scalar}

Scalar field dynamics is driven by the Klein Gordon 
equation. It is possible to define a gauge invariant 
form for the fluctuations in the expectation value 
of $\phi$ \cite{KS}, which we define as 
$\delta\phi_{GI}$; the form of the Klein Gordon equation 
for this quantities, in Fourier space at wavenumber $k$ is 
$$
\ddot{\delta\phi}_{GI}+2{\cal H}\dot{\delta\phi}_{GI}+
(k^{2}+a^{2}V_{\phi\phi})\delta\phi_{GI}=
$$
\begin{equation}
=\dot{\phi}(\dot{\Psi}-3\dot{\Phi})-
2a^{2}V_{\phi}\Psi\ ,
\label{KG}
\end{equation}
where $a$ is the cosmic scale factor, ${\cal H}=\dot{a}/a$ 
is the Hubble parameter, $\Psi$, $\Phi$ are the gauge invariant 
expressions for the scalar metric fluctuations, $()_{\phi}$ and 
$\dot{()}$ mean derivatives with respect to $\phi$ and conformal 
time respectively. The background reduces to a Cosmological 
Constant equivalent case when $V_{\phi}=0$: in this case 
$\phi =constant$ is a solution of the unperturbed Klein 
Gordon equation, reducing Eq.(\ref{KG}) to 
\begin{equation}
\ddot{\delta\phi}_{GI}+2{\cal H}\dot{\delta\phi}_{GI}+
(k^{2}+a^{2}V_{\phi\phi})\delta\phi_{GI}=0\ .
\label{KGmass}
\end{equation}
Therefore, even in conditions in which the background expectation 
value is frozen providing $w_{Q}=-1$ like a Cosmological Constant, 
some mechanism in the early universe \cite{PB} and/or some coupling 
with other fields entering in $V_{\phi\phi}$ \cite{EQ,CQ}
could excite fluctuations, providing an unique signature on 
the nature of dark energy.\\ 
In addition, it should be noted that if the equation of 
state is near but different from -1, as one has if $\phi$ 
rolls in an almost flat potential, Quintessence fluctuations are 
dragged to be non-zero by fluctuations in the gravitational 
potential itself, as the following simple argument shows. 
In absence of anisotropic stress, the relation 
$\Psi =-\Phi$ holds, and it is possible to show that in a matter 
or radiation dominate regimes $\Psi$ remain approximately constant 
\cite{HS}. Neglecting the term involving $\dot{\phi}$ 
and in the limit of slow expansion 
the right hand side of Eq. (\ref{KG}) gets constant 
and the simple solution is 
\begin{equation}
\delta\phi_{GI}\simeq -{2V_{\phi}\Psi\over k^{2}/a^{2}+V_{\phi\phi}}\ ,
\label{dragging}
\end{equation}
which means that metric fluctuations drag the corresponding 
ones in the scalar field. The behavior (\ref{dragging}) has 
been found rather independent on the details of initial 
conditions \cite{BRAX}.\\ 
The simple arguments exposed in this Section clarly show as 
the investigation of the nature of dark energy cannot be 
reduced to the measure of its equation of state, but deserves 
a deep understanding of all the possible physical degrees 
of freedom associated to this component. 
\begin{figure}
\includegraphics[height=2.5in,width=2.5in,angle=90]{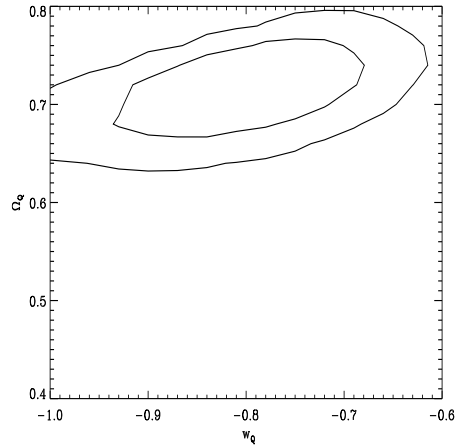}
\caption{$1\sigma$ and $2\sigma$ contours in the 
$(w_{Q},\Omega_{Q})$ plane. Cosmological Constant is 
recovered at 2$\sigma$.}
\label{f2}
\end{figure}

\section{Constraints from CMB}
\label{constraints}

According to the arguments exposed above, even in the case 
in which observations indicate $w_{Q}=-1$, this would not be 
a conclusive evidence in favour of a Cosmological Constant. 
However, the quest for $w_{Q}$ is a well posed problem, with solid 
theoretical motivations, already addressed by several authors 
\cite{BRAX,AME,OUR1,OUR2,MELCH,SVED}. Due to the large number of 
cosmological parameters, a conclusive statement is 
likely to be given by the planned and future experiments like SNAP, 
{\sc SNfactory}, MAP, {\sc Planck}, together with the data 
from redshift surveys. At the present, the only possible strategy is 
to fix some of the cosmological parameters at reasonable values. 
In \cite{OUR2} we assumed flat cosmologies with a fixed Hubble 
constant within the allowed range by the Hubble Space Telescope 
\cite{HST}. We allowed variations of baryon abundance, 
spectral index of scalar fluctuations, cosmological gravitational 
waves with spectral index and amplitude related by single field 
inflationary consistency relations; relaxing this 
hypotesis in particular can largely affect the results \cite{EFSTA}. \\ 
We compared the predicted CMB spectra in tracking scalar 
field cosmologies with the available data; the effects on 
the CMB in these scenarios are well known and have high 
amplitude, resulting mainly in a shift of the acoustic 
peaks toward small multipoles, as well as an increased 
Integrated Sachs Wolfe (ISW) effect on low multipoles 
as $w_{Q}$ increases from -1 toward larger values, 
as it is clear from figure \ref{f1}.\\
We found dark energy as the dominant cosmological component
$\Omega_{Q}=0.71^{+0.05}_{-0.04}$, with equation of state
$w_{Q}=-0.82^{+0.14}_{-0.11}$ ($68\%$ C.L.); the confidence 
region is shown in figure \ref{f2}, which also shows how 
ordinary Cosmological Constant is recovered at 2$\sigma$. 
We argued that this intriguing result is due to a mild evidence of 
a projection effect in the available CMB data \cite{OUR2}. 
The best fit value of the physical baryon density is in good 
agreement with the primordial nucleosynthesis bound. Also, 
we find no significant evidence for deviations from scale-invariance, 
although a scalar spectral index slightly smaller than unity is 
marginally preferred. Finally, we find that the contribution of 
cosmological gravitational waves is negligible within our hypothesis.\\ 
In conclusion, within our assumptions we found a mild evidence in favour 
of a dark energy equation of state larger than -1. 
Incoming data and cross-correlation with other unbiased 
observations can help to further check this intriguing result.

\end{document}